\documentclass[letterpaper]{article} 
\usepackage{snnet}  
\usepackage{times}  
\usepackage{helvet} 
\usepackage{courier}  
\usepackage[hyphens]{url}  
\usepackage{graphicx} 
\usepackage{graphicx}  
\usepackage{natbib}  
\usepackage{caption} 
\usepackage{amsfonts}  
\usepackage{amsmath}  
\usepackage{multirow} 
\usepackage{tablefootnote}  
\usepackage[switch]{lineno}
\nocopyright
\title{Interactive Speech and Noise Modeling for Speech Enhancement}
\author {
        Chengyu Zheng\textsuperscript{\rm 1}\thanks{The work was done at Microsoft Research Asia.},
        Xiulian Peng\textsuperscript{\rm 2},
        Yuan Zhang\textsuperscript{\rm 1},
        Sriram Srinivasan\textsuperscript{\rm 3},
        Yan Lu \textsuperscript{\rm 2} \\
}
\affiliations {
    \textsuperscript{\rm 1} Communication University of China\\ 
    \textsuperscript{\rm 2} Microsoft Research Asia\\ 
    \textsuperscript{\rm 3} Microsoft Corporation\\ 
    zhengchengyu@cuc.edu.cn, xipe@microsoft.com, yzhang@cuc.edu.cn, Sriram.Srinivasan@microsoft.com, yanlu@microsoft.com
}

\begin{document}
\maketitle
\begin{abstract}
    Speech enhancement is challenging because of the diversity of background noise types. Most of the existing methods are focused on modelling the speech rather than the noise. In this paper, we propose a novel idea to model speech and noise simultaneously in a two-branch convolutional neural network, namely SN-Net. In SN-Net, the two branches predict speech and noise, respectively. Instead of information fusion only at the final output layer, interaction modules are introduced at several intermediate feature domains between the two branches to benefit each other. Such an interaction can leverage features learned from one branch to counteract the undesired part and restore the missing component of the other and thus enhance their discrimination capabilities. We also design a feature extraction module, namely residual-convolution-and-attention (RA), to capture the correlations along temporal and frequency dimensions for both the speech and the noises. Evaluations on public datasets show that the interaction module plays a key role in simultaneous modeling and the SN-Net outperforms the state-of-the-art by a large margin on various evaluation metrics. The proposed SN-Net also shows superior performance for speaker separation. 
\end{abstract}

\section{Introduction}
\begin{figure}[t]
    \centering
    \includegraphics[width=0.47\textwidth]{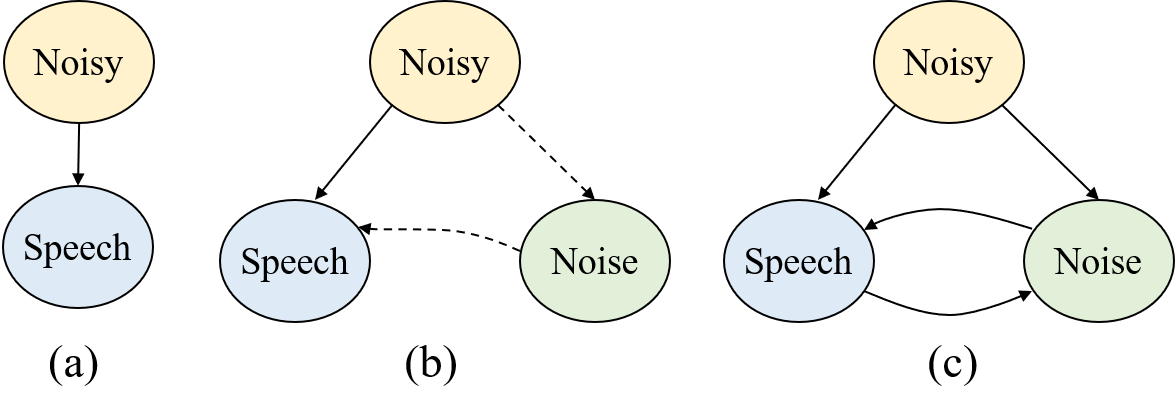}
    \caption{Illustration of different methods. (a) Most existing deep-learning-based methods directly model speech. (b) Most traditional methods predict speech with noise estimate. (c) Our method simultaneously models speech and noise with information interaction.}
    \label{fig0}
\end{figure}

\noindent Speech enhancement aims at separating speech from background interference signals. Mainstream deep learning-based methods learn to predict the speech signal in a supervised manner, as shown in Figure \ref{fig0} (a). Most prior works operate in the time-frequency (T-F) domain by predicting a mask between noisy and clean spectra \cite{wang2014IRM, williamson2015complex} or directly predicting the clean spectrum \cite{xu2013experimental, tan2018convolutional}. Some methods operate in the time domain by estimating speech signals from raw-waveform noisy signals in an end-to-end way \cite{fu2017raw, pascual2017segan, pandey2019tcnn}. These methods have considerably improved the quality of enhanced speech compared with traditional signal processing based schemes. However, speech distortion or residual noise can often be observed in the enhanced speech, showing that there are still correlations between predicted speech and the residual signal by subtracting enhanced speech from noisy signal.

\begin{figure*}[t]
    \centering
    \includegraphics[width=0.95\textwidth]{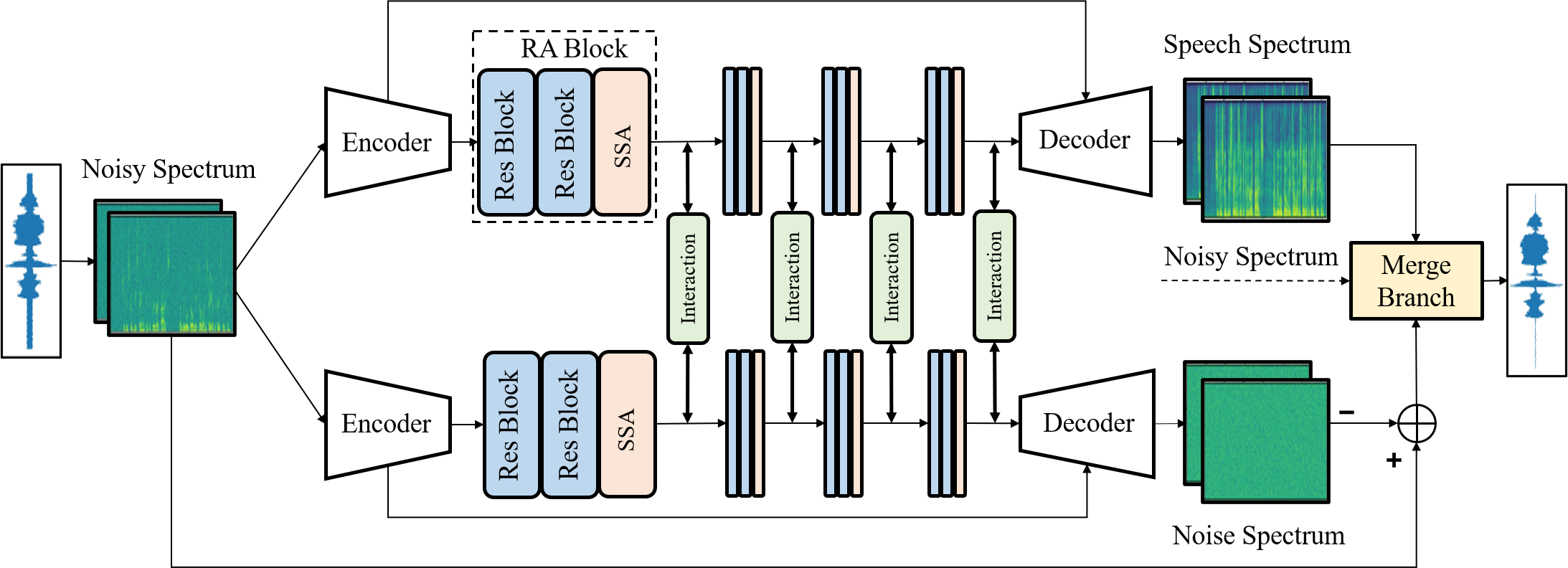} 
    \caption{Overall network structure of SN-Net.}
    \label{fig1}
\end{figure*}

Instead of only predicting speech and ignoring the characteristics of background noises, traditional signal processing and modeling based methods mostly take the other way (see Figure \ref{fig0} (b)), i.e. estimating noise or building noise models for speech enhancement \cite{boll1979suppression, hendriks2010noisePSD, wang2017model, wilson2008NMF1, moham2013NMF2}. Some model-based methods instead model both speech and noise \cite{srinivasan2006Codebook, srinivasan2005Codebook}, possibly with alternate model update. However, they typically cannot generalize well when prior noise assumption cannot be met or the interference signal is not structured. In deep-learning-based methods, two recent attempts \cite{odelowo2017noise, odelowo2018study} focus on directly predicting noise considering that noise is dominant in low-SNR conditions. However, the benefit is limited.


The remaining correlation between predicted speech and noise motivates us to explore the information flow between speech and noise estimations, as shown in Figure \ref{fig0} (c). Since speech-related information exists in predicted noise, and vice versa, adding information communication between them may help to recover some missing components and remove undesired information from each other. In this paper, we propose a two-branch convolutional neural network, namely SN-Net, to simultaneously predict speech and noise signals. Between them are information interaction modules, by which noise or speech related information are extracted from the noise branch and added back to speech features to counteract the undesired noise part or recover the missing speech, and vice versa. In this way, the discrimination capability is largely enhanced. The two branches share the same network structure, which is an encoder-decoder-based model with several residual-convolution-and-attention (RA) blocks in between for separation. Motivated by the success of self-attention technique in machine translation and computer vision tasks \cite{vaswani2017attention, wang2018nonlocal}, we propose to combine temporal self-attention and frequency-wise self-attention parallelly inside each RA block for capturing global dependency along temporal and frequency dimensions in a separable way.

Our main contributions are summarized as follows.

\begin{itemize}
    \item We propose to simultaneously model speech and noise in a two-branch deep neural network and introduce information flow between them. In this way, speech part is enhanced while residual noise is suppressed for speech estimation, and vice versa.
    \item We propose a RA block for feature extraction. Separable self-attention is utilized in this block to globally capture the temporal and frequency dependencies.
    \item We validate the superiority of proposed scheme in an ablation study and comparison with state-of-the-art algorithms on two public datasets. Moreover, we extend our method to speaker separation, which also shows great performance. These results demonstrate the superiority and potential of the proposed method.
\end{itemize}

\section{Related Work}

\begin{figure*}[ht]
    \centering
    \includegraphics[width=0.5\textwidth]{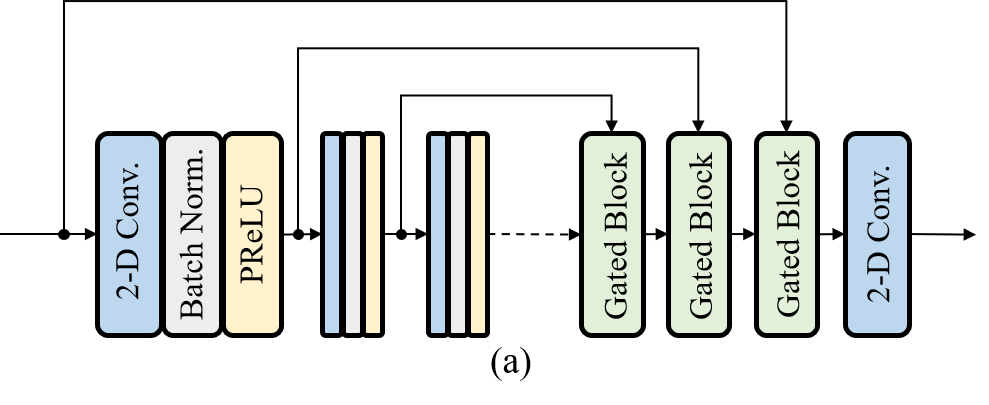}
    \includegraphics[width=0.4\textwidth]{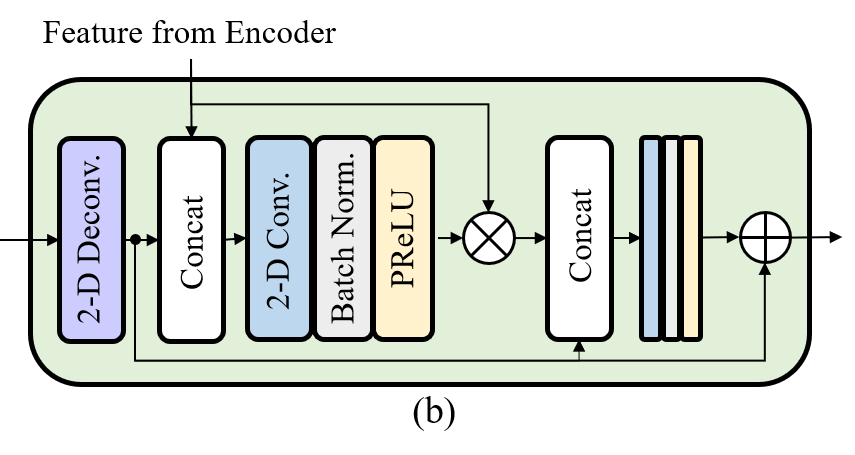}
    \caption{(a) Encoder-decoder structure. The dashed arrow denotes the separation module using RA blocks. (b) Detailed structure of the gated block inside the decoder.}
    \label{fig2}
\end{figure*}

\subsection{Deep Learning-based Speech Enhancement}

\noindent Deep learning-based methods mainly study how to build a speech model. According to the adopted signal domain, these methods can be classified into two categories. Time-Frequency (T-F) domain methods take T-F representation, either complex or log power spectrum of the magnitude, as input. They typically estimate a real or complex ratio mask for each T-F bin to map noisy spectra to speech spectra \cite{williamson2015complex, wang2014IRM, choi2019phase} or directly predict the speech representation \cite{xu2013experimental, tan2018convolutional}. Time-domain methods take waveform as input and typically extract a hidden representation of the raw waveform through an encoder and reconstruct an enhanced version from that \cite{fu2017raw, pascual2017segan, pandey2019tcnn}. Although these methods have shown great improvements over traditional methods, they only focus on modeling speech and neglect the importance of understanding noise characteristics.

\subsection{Noise-Aware Speech Enhancement}

\noindent Noise information is often considered in traditional signal processing based methods \cite{boll1979suppression, hendriks2010noisePSD, wang2017model} with prior distribution assumptions for speech and noise. However, it is a challenging task to estimate the noise power spectral density for non-stationary noises and thus mostly stationary noise is assumed. They are unsuitable in generalization to low SNR and non-stationary noise conditions. Instead, some model-based methods build models for speech and noise and show more promising results, e.g., codebook \cite{srinivasan2006Codebook, srinivasan2005Codebook} and nonnegative matrix factorization (NMF) \cite{wilson2008NMF1, moham2013NMF2} based methods. However, they either need prior knowledge of the noise type \cite{srinivasan2006Codebook, srinivasan2005Codebook} or are only effective for structured noise \cite{wilson2008NMF1, moham2013NMF2}; therefore their generalization capability is limited.

Deep learning-based methods can better generalize to various noise conditions. There are also some attempts on incorporating noise information, for example, by adding constraints to loss functions \cite{fan2019noise, xu2020using, xia2020weighted} or by directly predicting noise instead of speech \cite{odelowo2017noise, odelowo2018study}. The former does not model noise at all and the characteristics of noise are not exploited. The latter loses the speech information and show even worse quality than corresponding speech prediction method in low SNR and unseen noise conditions. A more relevant work utilizes two deep auto encoders (DAEs) to estimate speech and noise \cite{sun2015unseen} . It first trains a DAE for speech spectrum reconstruction and then introduces another DAE to model noise with the constraint that the sum of outputs of the two DAEs is equal to the noisy spectrum.

Different from aforementioned approaches, we proposed a two-branch CNN to predict speech and noise simultaneously and introduce interaction modules at several intermediate layers to make them benefit from each other. Such a paradigm makes it suitable for speaker separation as well. 

\subsection{Two-Branch Neural Networks}

\noindent Two-branch neural networks have been explored in various tasks for capturing cross-modality information \cite{nam2017dual, image-text} or different levels of information \cite{simonyan2014two, wang2020dual}. For speech enhancement, a two-branch modeling is proposed to predict the amplitude and phase of the enhanced signal, respectively \cite{yin2020phasen}. In this paper, we aim to exploit the two correlated tasks, i.e. speech and noise estimations and explicitly modeling them in an interactive two-branch framework for better discrimination.

\subsection{Self-Attention Model}

\noindent Self-attention mechanism has been widely used in many tasks, e.g., machine translation \cite{vaswani2017attention}, image generation \cite{zhang2019self} and video question answering \cite{li2019beyond}. For video, spatio-temporal attention is also considered to exploit long-term dependency along both spatial and temporal dimensions \cite{wu2019video}. Recently, speech-related tasks have also benefited from self-attention, e.g., speech recognition \cite{salazar2019ctc} and speech enhancement \cite{kim2020tgsa, koizumi2020speech}. In these works, self-attention is applied along the temporal dimension only, neglecting the global dependency inside each frame. Motivated by the spatio-temporal attention in video-related tasks, we propose to employ both frequency-wise and temporal self-attention to better capture dependencies along different dimensions. Such an attention is employed in both speech and noise branches for simultaneous modeling the two signals.

\section{Proposed Method}

\subsection{Overview}

\noindent Figure \ref{fig1} shows the overall network structure of SN-Net. The input is the complex T-F spectrum computed by short-time Fourier transform (STFT), denoted as $ X^I\in R^{T\times F\times 2} $, where $ T $ is the number of frames and $ F $ is the number of frequency bins. There are two branches in SN-Net, one of which predicts speech and the other predicts noise. They share the same network structure but have separate network parameters. Each branch is an encoder-decoder based structure, with several RA blocks inserted inbetween them. In this way, it is capable of simultaneously mining the potential of different components of the noisy signal. Between the two branches are interaction modules designed to transform and share information. After each branch gets its output, a merge branch is employed to adaptively combine the two outputs to generate the final enhanced speech.

\subsection{Encoder and Decoder}

\noindent As shown in Figure \ref{fig2} (a), the encoder has three 2-D convolutional layers, each with a kernel size of (3, 5). The stride is (1, 1) for the first layer and (1, 2) for the following two. The channel numbers are 16, 32, 64, respectively. As a result, the output feature of the encoder is $ \mathcal{F}^E_k\in \mathbb{R}^{T\times F'\times C} $, where $ F'=\frac{F}{4}$, $C=64 $ and $ k\in \left\{S,N\right\} $. $ S $ and $ N $ denote speech and noise branches, respectively. For simplicity, the subscript $ k $ will be ignored in the following.

The decoder consists of three gated blocks followed by one 2-D convolutional layer, which reconstructs the output $ \mathcal{F}^D\in \mathbb{R}^{T\times F\times 2} $. As shown in Figure \ref{fig2} (b), the gated block learns a multiplicative mask on corresponding feature from the encoder, aiming to suppress its undesired part. The masked encoder feature is then concatenated with the deconvolutional feature and fed into another 2-D convolutional layer to generate the residual representation. After three gated blocks, the final convolutional layer learns the amplitude gain and the phase for reconstruction, similar to that in \cite{choi2019phase}. The kernel size for all 2-D deconvolutional layers is (3,5). The stride is (1,2) for the first two gated blocks and (1,1) for the last one. The channel numbers are 32, 16, 2, respectively. All the 2-D convolutional layers in the decoder have a kernel size of (1,1), a stride of (1,1) and a channel number the same as that of their deconv layers.

All the convolutional layers in the encoder and the decoder are followed by a batch normalization (BN) and a parametric ReLU (PReLU). No down-sampling is performed along the temporal dimension to preserve the temporal resolution.

\subsection{RA Block}

\begin{figure}[t]
    \centering
    \includegraphics[width=0.3\textwidth]{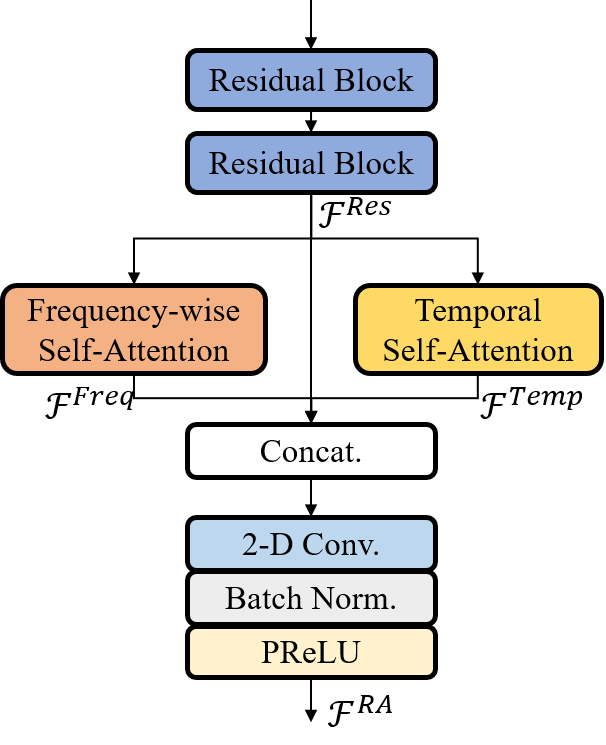}
    \caption{Structure of the RA block.}
    \label{fig3}
\end{figure}

\noindent The RA block is designed to extract features and perform separation for both speech and noise branches. It is challenging because of the diversities of noise types and the difference between speech and noises. We employ the separable self-attention (SSA) technique to capture the global dependencies along temporal and frequency dimensions, respectively. It is intuitive to use attention for these two dimensions as humans tend to put more attention to some parts of an audio signal (e.g., speech) while less to the surrounding part (e.g., noise) and they perceive differently on different frequencies. When it comes to the speech-noise network in SN-Net, the SSA modules in speech and noise branches perceive signals differently, which will be demonstrated in the ablation study section afterwards. 

In SN-Net, there are four RA blocks between the encoder and the decoder. Each block consists of two residual blocks and a SSA module, as shown in Figure \ref{fig3}, capturing both local and global dependencies inside the signal. Each residual block has two 2-D convolutional layers with a kernel size of (5,7), a stride of (1,1) and the same number of channels as their inputs. The output feature of two residual blocks $ \mathcal{F}^{Res}_i\in \mathbb{R}^{T\times F’\times C} $ ($ i\in \left\{1,2,3,4\right\} $ represents the $ i^{th} $ RA block and will be ignored in the following) is fed parallelly into temporal self-attention and frequency-wise self-attention blocks. These two attention blocks produce the outputs $ \mathcal{F}^{Temp}\in \mathbb{R}^{T\times F’\times C} $ and $ \mathcal{F}^{Freq}\in \mathbb{R}^{T\times F’\times C} $. The three features $\mathcal{F}^{Res} $, $ \mathcal{F}^{Temp} $ and $ \mathcal{F}^{Freq} $ are then concatenated and fed into a 2-D convolutional layer to generate the block output $ \mathcal{F}^{RA}\in \mathbb{R}^{T\times F’\times C} $, used in the interaction module.

For self-attention, we employ the scaled dot-product self-attention here. Considering the computational complexity, channels are reduced by half inside SSA. The temporal self-attention can be represented as
\begin{equation}
\begin{aligned} 
    \mathcal{F}_t^{k}&=Reshape^t(Conv(\mathcal{F}^{Res})), k\in \left\{K, Q, V\right\},\\
    SA^{t}&=Softmax(\mathcal{F}_t^{Q}\cdot (\mathcal{F}_t^{K})^T/\sqrt{\frac{C}{2}\times F'})\cdot \mathcal{F}_t^{V},\\
    \mathcal{F}^{Temp}&=\mathcal{F}^{Res}+Conv(Reshape^{t*}(SA^{t})),
\end{aligned}
\end{equation}
where $\mathcal{F}_t^{k} \in \mathbb{R}^{T\times (\frac{C}{2}\times F')}$, $SA^{t}\in \mathbb{R}^{T\times (\frac{C}{2}\times F')}$ and $\mathcal{F}^{Temp} \in \mathbb{R}^{T\times F'\times C}$, respectively. $(\cdot)$ denotes matrix multiplication. $Reshape^t(\cdot)$ denotes a tensor reshape from $\mathbb{R}^{T\times F'\times \frac{C}{2}}$ to $\mathbb{R}^{T\times (\frac{C}{2}\times F')}$ and $Reshape^{t*}(\cdot)$ is the opposite. The frequency-wise self-attention is given by
\begin{equation}
\begin{aligned} 
    \mathcal{F}_f^{k}&=Reshape^f(Conv(\mathcal{F}^{Res})), k\in \left\{K, Q, V\right\},\\
    SA^{f}&=Softmax(\mathcal{F}_f^{Q}\cdot (\mathcal{F}_f^{K})^T/\sqrt{\frac{C}{2}\times T})\cdot \mathcal{F}_f^{V},\\
    \mathcal{F}^{Freq}&=\mathcal{F}^{Res}+Conv(Reshape^{f*}(SA^{f})),
\end{aligned}
\end{equation}
where $\mathcal{F}_f^{k}\in \mathbb{R}^{F'\times (\frac{C}{2}\times T)}$, $SA^{f}\in \mathbb{R}^{F'\times(\frac{C}{2}\times T)}$ and $\mathcal{F}^{Freq}\in \mathbb{R}^{T\times F'\times C}$, respectively. $Reshape^f(\cdot)$ reshapes a tensor from $\mathbb{R}^{T\times F'\times \frac{C}{2}}$ to $\mathbb{R}^{F'\times (\frac{C}{2}\times T)}$.

\begin{figure}[t]
    \centering
    \includegraphics[width=0.33\textwidth]{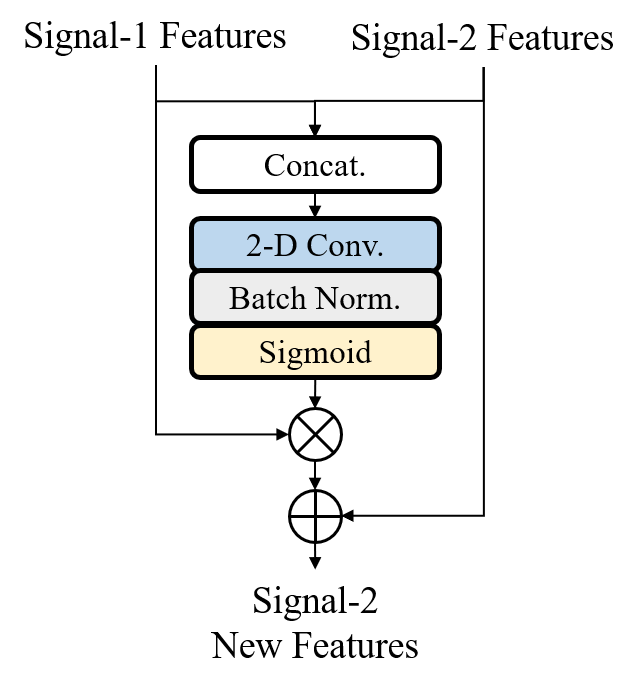}
    \caption{Structure of the interaction module.}
    \label{fig4}
\end{figure}

In the above equations, $ Conv $ denotes a convolutional layer followed by BN and PReLU. All the convolutional layers have a kernel size of (1,1) and a stride of (1,1). 

\subsection{Interaction Module}

\noindent In SN-Net, the speech and noise branches share the same input signal, which suggests that the internal features of two branches are correlated. In light of this, we propose an interaction module to exchange information between the branches. With this block, information transformed from the noise branch is expected to enhance the speech part and counteract the noise features inside the speech branch, and vice versa. We will show in ablation study afterwards that this module plays a key role in simultaneously modeling the speech and noises. 

The structure of the interaction module is shown in Figure \ref{fig4}. Taking speech branch as an example, feature from the noise branch $ \mathcal{F}_N^{RA} $ is first concatenated with that from the speech branch $ \mathcal{F}_S^{RA}$. They are then fed into a 2-D convolutional layer to generate a multiplicative mask $ \mathcal{M}^N $, predicting the suppressed and preserved areas of $ \mathcal{F}_N^{RA} $. A residual representation $ \mathcal{H}^{N2S} $ is then obtained by multiplying $ \mathcal{M}^N $ with $ \mathcal{F}_N^{RA} $ elementally. Finally, the block adds $ \mathcal{F}_S^{RA} $ and $ \mathcal{H}^{N2S} $ to get a “filtered” version of the speech feature, which will be fed into the next RA block. The process is given by
\begin{equation}
\begin{aligned} 
    \mathcal{F}_{S_{out}}^{RA}&=\mathcal{F}_S^{RA}+\mathcal{F}_N^{RA}*Mask(\mathcal{F}_N^{RA},\mathcal{F}_S^{RA}),\\
    \mathcal{F}_{N_{out}}^{RA}&=\mathcal{F}_N^{RA}+\mathcal{F}_S^{RA}*Mask(\mathcal{F}_S^{RA},\mathcal{F}_N^{RA}),
\end{aligned}
\end{equation}
where $Mask(\cdot)$ is short for concatenation, convolution and sigmoid operations. $(*)$ denotes element-wise multiplication.

\subsection{Merge Branch}

\noindent After reconstructing the speech and noise signals in two branches, a merge module is further employed to combine the two outputs. This is done in the time domain to achieve the cross-domain benefit \cite{Kim2018MDPhD}. The two decoder outputs are transformed to time-domain and overlapped framed representation using the same window length as the STFT we use, resulting in $ \tilde{s}\in \mathbb{R}^{T\times K} $ and $ \tilde{n}\in \mathbb{R}^{T\times K} $, where $K$ is the frame size. These two representations are stacked with the noisy waveform $ x $ and fed into the merge branch. The merge network uses a 2-D convolutional layer, followed by an temporal self-attention block to capture global temporal dependency and two other convolutional layers to learn an element-wise mask $ m\in \mathbb{R}^{T\times K} $. The kernel size of all three convolutional layers is (3,7) and the channel number is 3, 3, 1, respectively. BN and PReLU are used after each convolutional layer except the last one. Sigmoid activation is used in the last layer. Finally, the 2D enhanced signal is obtained by
\begin{equation}
    \hat{s} = m\times \tilde{s} + (1-m)\times (x-\tilde{n}).
\end{equation}
The 1D signal is reconstructed from $\hat{s}$ after overlap and add.

\section{Experiments}

\subsection{Datasets}

\noindent Three public datasets are used in our experiments.

\textbf{DNS Challenge} The DNS challenge \cite{reddy2020interspeech} at Interspeech 2020 provides a large dataset for training. It includes 500 hours clean speech across 2150 speakers collected from Librivox and 60000 noise clips from Audioset \cite{gemmeke2017audio} and Freesound with 150 classes. For training, we synthesized 500 hours noisy samples with SNR levels of -5dB, 0dB, 5dB, 10dB and 15dB. For evaluation, we use 150 synthetic noisy samples without reverberation inside the test set, whose SNR levels are randomly distributed between 0 dB and 20 dB.

\textbf{Voice Bank + DEMAND} This is a small dataset created by Valentini-Botinhao et al. \cite{valentini2016investigating}. Clean speech clips are collected from the Voice Bank corpus \cite{veaux2013voice} with 28 speakers for training and another 2 unseen speakers for test. Ten noise types with two artificially generated and eight real recordings from DEMAND \cite{thiemann2013diverse} are used for training. Five other noise types from DEMAND are chosen for the test, without overlapping with the training set. The SNR values are 0dB, 5dB, 15dB and 20dB for training and 2.5dB, 7.5dB, 12.5dB and 17.5dB for test. 

\textbf{TIMIT Corpus} This dataset is used for our speaker separation experiment. It contains recordings of 630 speakers, each reading 10 sentences and there are 462 speakers in the training set and 168 speakers in the test set. Two sentences from different speakers are mixed with random SNRs to generate mixture utterances. Shorter sentences are zero padded to match the size of longer ones. In total, the training set includes 4620 sentences and the test set 1680 sentences.

\begin{table}[t]
    \centering
    \begin{tabular}{lrcc}
        \hline
        \multicolumn{2}{l}{\textbf{Models}} & \textbf{SDR(dB)} & \textbf{PESQ}\\
        \hline
        \multicolumn{2}{l}{Noisy} & 9.09 & 1.58\\
        \hline
        \multicolumn{2}{l}{Speech branch w/o SSA (baseline)} & 18.06 & 3.05\\
        \multicolumn{2}{l}{Speech branch} & 18.75 & 3.28\\
        \hline
        \multicolumn{2}{l}{SN-Net w/o interaction} & 19.04 & 3.29 \\
        \multicolumn{2}{l}{SN-Net} & \textbf{19.52} & \textbf{3.39} \\
        \hline
    \end{tabular}
    \caption{Ablation study on DNS Challenge dataset}
    \label{table1}
\end{table}

\subsection{Evaluation Metrics}

\noindent To evaluate the quality of the enhanced speech, the following objective measures are used. Higher scores indicate better quality.

\begin{itemize}
    \item SSNR: Segmental SNR.
    \item SDR \cite{vincent2006performance}: Signal-to-distortion ratio.
    \item PESQ \cite{rec2005p}: Perceptual evaluation of speech quality, using the wide-band version recommended in ITU-T P.862.2 (from -0.5 to 4.5).
    \item CSIG \cite{hu2007evaluation}: Mean opinion score (MOS) prediction of the signal distortion (from 1 to 5).
    \item CBAK \cite{hu2007evaluation}: MOS prediction of the intrusiveness of background noises (from 1 to 5).
    \item COVL \cite{hu2007evaluation}: MOS prediction of the overall effect (from 1 to 5).
\end{itemize}

\subsection{Implementation Details}
\subsubsection{Input}
All signals are resampled to 16kHz and clipped to 2 seconds long. We take the STFT complex spectrum as input, with a Hann window of length 20ms, a hop length of 10ms and a DFT length of 320. 

\subsubsection{Loss Function}
The loss function includes three terms, i.e. $ \mathcal{L}= \mathcal{L}_{Speech}+\alpha\mathcal{L}_{Noise}+\beta\mathcal{L}_{Merge} $, where $ \mathcal{L}_{Speech} $, $ \mathcal{L}_{Noise} $ and $ \mathcal{L}_{Merge}$ represent the loss of three branches, respectively. $ \alpha$ and $\beta $ are weighting factors balancing among the three. All terms use a mean-squre-error (MSE) loss on the power-law compressed STFT spectrum \cite{ephrat2018looking}. An inverse STFT and forward STFT are conducted on speech and noise branches before calculating the loss to ensure STFT consistency as that in \cite{wisdom2019differentiable}.

\subsubsection{Training}
The proposed algorithm is implemented in TensorFlow. We use adam optimizer with a learning rate of 0.0002. All the layers are initialized with Xavier initialization. The training is conducted in two stages. The speech and noise branches are jointly trained first with the loss weight $ \alpha=1 $ and $ \beta=0 $. Then the merge branch is trained with the parameters of previous two fixed, using only the loss $\mathcal{L}_{Merge}$. We train both stages for 60 epochs for DNS Challenge and 400 epochs for Voice Bank + DEMAND dataset. The batch size for all experiments is set to 32, unless otherwise specified. 

\begin{figure}[t]
    \centering
    \includegraphics[width=0.48\textwidth]{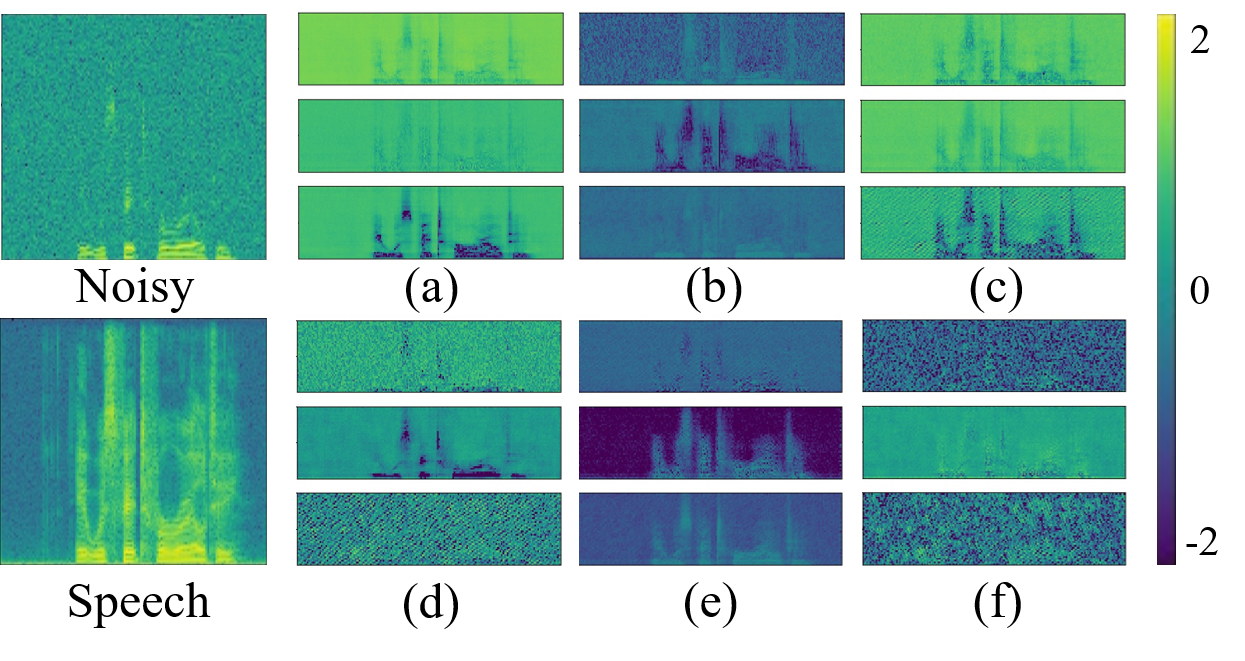}
    \caption{Log-scale feature visualization for the fourth interaction module. (a) Input feature of speech branch. (b) Transformed feature from noise to speech branch. (c) Output feature of speech branch. (d) Input feature of noise branch. (e) Transformed feature from speech to noise branch. (f) Output feature of noise branch. Three channels with the highest activities are visualized here.}
    \label{fig7}
\end{figure}

\subsection{Ablation Study}
\subsubsection{Objective Quality}
We first evaluate the effectiveness of different parts of the proposed SN-Net based on the DNS Challenge dataset. As shown in Table \ref{table1}, we take the speech branch without SSA as the baseline. After adding SSA to the single-branch model, we observe a 0.69 dB gain on SDR and 0.23 on PESQ. By comparing “Speech branch” with “SN-Net w/o interaction”, we can see that when no interaction is employed, adding another branch with merge module at the output only marginally improves the SDR by 0.29 dB and no improvement on PESQ. After introducing the information flow, it evidently improves the SDR by 0.77 dB and PESQ by 0.11 compared to single branch. These results verify the effectiveness of the proposed RA and interaction modules for simultaneously modeling speech and noises.

\begin{figure}[t]
    \centering
    \includegraphics[width=0.49\textwidth]{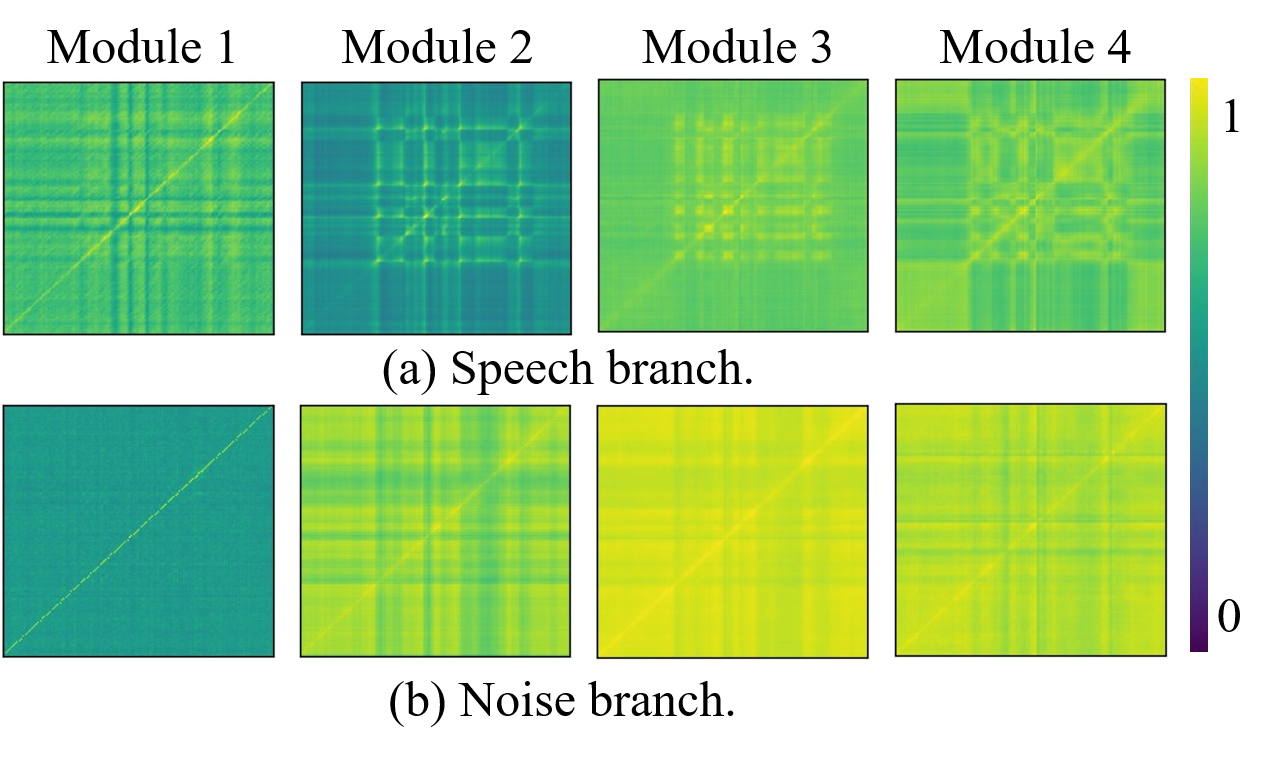}
    \caption{Visualization of temporal self-attention matrices from different RA blocks. (a) Speech branch. (b) Noise branch. Each matrix is linearly scaled to $[0, 1]$.}.
    \label{fig5}
\end{figure}

\begin{figure}[t]
    \centering
    \includegraphics[width=0.49\textwidth]{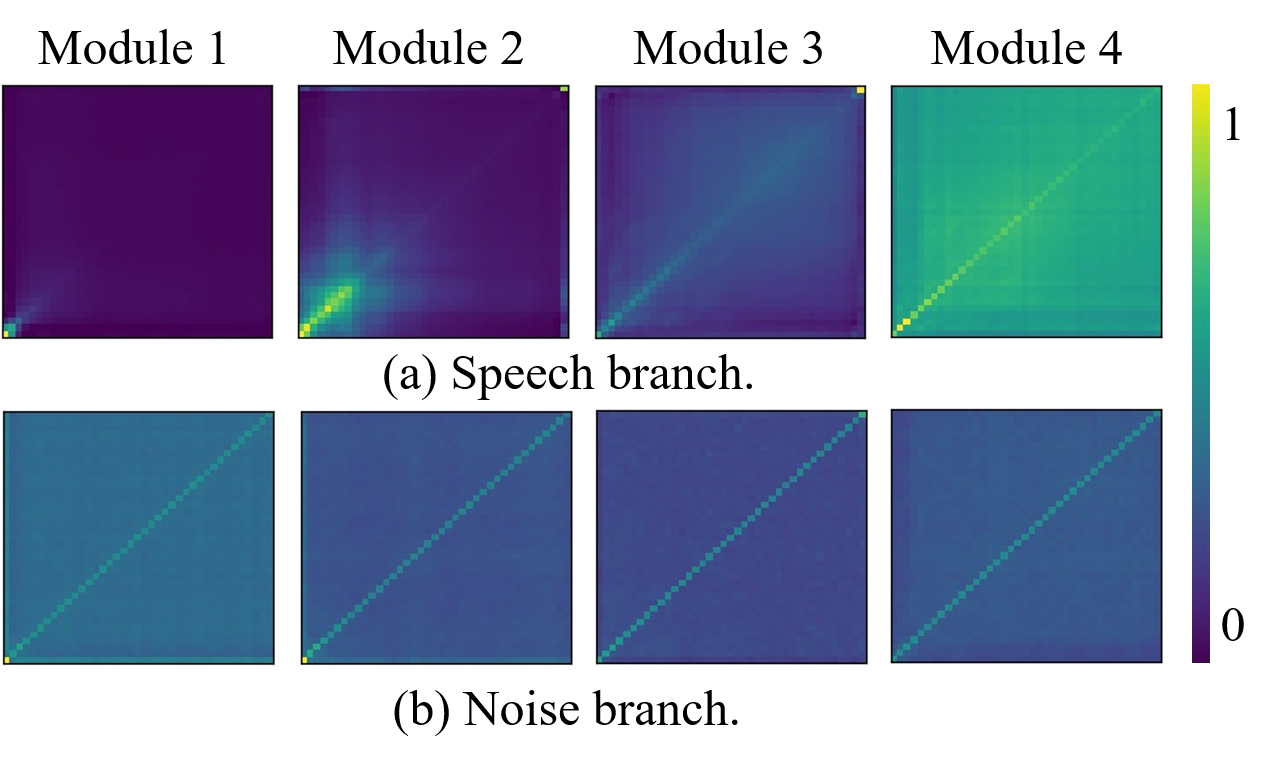}
    \caption{Visualization of frequency-wise self-attention matrices from different RA blocks. (a) Speech branch. (b) Noise branch. Each matrix is linearly scaled to $[0, 1]$.}
    \label{fig6}
\end{figure}

\subsubsection{Visualization of Information Flow}
In order to further understand how the interaction module works, we visualize the input feature, the output and the feature transformed from the other branch of this module in Figure \ref{fig7}. An audio signal corrupted by white noises is used for illustration, whose spectrum is shown in the first column.

The transformed feature shown in Figure \ref{fig7} (b) is learned from the feature in (d) and added to the feature in (a), resulting in the output feature of speech branch in (c) and vice versa. Comparing (a) and (c), we can see that the speech area is better separated with noise after interaction. For noise branch, the speech part is mostly removed in (f) compared with (d). These results show that the interaction module indeed helps the simultaneous speech and noise modeling with better separation capabilities. In terms of interchanged information, the undesired speech part in (d) is counteracted by features learned from the speech branch (e.g., the second channel of the noise branch) and the undesired noise part in (a) is suppressed by features learned from the noise branch (e.g., the third channel of the speech branch). These observations comply with our previous analysis.

\subsubsection{Visualization of Separable Self-Attention}
We further visualize the attention matrix to explore what it has learned. Figure \ref{fig5} shows the temporal self-attention matrix inside different RA blocks for the same audio signal as that in Figure \ref{fig7}. From (a) and (b), we can see that besides the diagonal line, each frame shows strong attentiveness to other frames and speech and noise branches behave differently for each RA module. This is reasonable as the two branches model different signals and their focus differs. For noise branch, the attention goes from local to global as the network goes deeper. The noise branch shows wider attentiveness than the speech branch as white noises spread in all frames while speech signal occurs only at some time.

Figure \ref{fig6} shows the frequency-wise self-attention matrix for the same audio signal. For speech branch, the focus goes from low-frequency area to full frequencies and from local to global, showing that as the network goes deeper, the frequency-wise self-attention tends to capture global dependency along the frequency dimension. For noise branch, all four RA blocks show a local attention as white noises have a constant power spectral density.

\subsection{Comparison with the State-of-the-Art}

\subsubsection{Speech Enhancement}
\begin{table}[t]
    \centering
    \small
    \begin{tabular}{lccccc}
        \hline
        \textbf{Methods} & \textbf{SSNR} & \textbf{PESQ} & \textbf{CSIG} & \textbf{CBAK} & \textbf{COVL}\\
        \hline
        Noisy & 1.68 & 1.97 & 3.35 & 2.44 & 2.63\\
        \hline
        SEGAN & 7.73 & 2.16 & 3.48 & 2.94 & 2.80\\
        MMSE-GAN & - & 2.53 & 3.80 & 3.12 & 3.14\\
        \hline
        PHASEN & \textbf{10.18} & 2.99 & 4.21 & 3.55 & 3.62\\
        \hline
        Koizumi et al. & - & 2.99 & 4.15 & 3.42 & 3.57\\
        \hline
        Ours & 9.83 & \textbf{3.12} & \textbf{4.39} & \textbf{3.60} & \textbf{3.77} \\
        \hline
    \end{tabular}
    \caption{Quality comparisons on Voice Bank + DEMAND}
    \label{table2}
\end{table}

\begin{table}[t]
    \centering
    \begin{tabular}{lccc}
        \hline
        \textbf{Methods} & \textbf{SDR(dB)} & \textbf{PESQ} \\
        \hline
        Noisy & 9.09 & 1.58 \\
        \hline
        TCNN & 16.86 & 2.34  \\
        TCNN-L & 16.58 & 2.78 \\
        \hline
        Conv-TasNet-SNR & - & 2.73 \\
        \hline
        DTLN & 16.54 & 2.34 \\
        \hline
        MultiScale+ & - & 2.71 \\
        \hline
        PoCoNet & - & 2.75 \\
        \hline
        Ours & \textbf{19.52} & \textbf{3.39} \\
        \hline
    \end{tabular}
    \caption{Quality comparisons on DNS Challenge}
    \label{table3}
\end{table}

\noindent Table \ref{table2} shows the comparisons with state-of-the-art methods on Voice Bank + DEMAND. SEGAN \cite{pascual2017segan} and MMSE-GAN \cite{soni2018time} are two GAN-based methods. PHASEN \cite{yin2020phasen} is a two-branch T-F domain approach where one branch predicts the amplitude and the other predicts the phase. Koizumi et al. \cite{koizumi2020speech} is a multi-head self-attention based method. Our method outperforms all of them in almost all metrics. The large improvements on PESQ, CSIG and COVL indicate that our method preserves better speech quality. 

Table \ref{table3} shows the comparison with state-of-the-art methods on DNS Challenge dataset. TCNN \cite{pandey2019tcnn} is a time-domain low-latency approach. We implemented two versions of it.  “TCNN” is exactly the same as described in the paper and “TCNN-L” is the long-latency version using the same T-F domain loss function as ours. Conv-TasNet-SNR \cite{koyama2020exploring} and DTLN \cite{westhausen2020dual} are real-time approaches. MultiScale+ \cite{choi2020phase} and PoCoNet \cite{isik2020poconet} are non-real-time methods, among which the PoCoNet took 1st place in the 2020 DNS challenge's Non-Real-Time track. Since narrow-band PESQ number was reported in the DTLN paper, we used the released model\footnote{https://github.com/breizhn/DTLN} to generate the enhanced speech and compute the metrics. For other methods, we use the numbers reported in their papers. Our method outperforms all of them by a large margin.

\begin{table}[t]
    \centering
    \begin{tabular}{lcc}
        \hline
        \textbf{Methods} & \textbf{SDRi(dB)} & \textbf{PESQ}\tablefootnote{Note that Conv-TasNet outputs 8khz audios. We use narrow-band PESQ here instead of wide-band. Accordingly, we downsample audios to 8khz for our method to match this evaluation.}\\
        \hline
        Conv-TasNet & 7.57 & 2.14\\
        Ours & \textbf{8.39} & \textbf{2.50} \\
        \hline
    \end{tabular}
    \caption{Two-speaker speech separation on TIMIT}
    \label{table4}
\end{table}

\subsubsection{Extension to Speaker Separation}

\noindent As SN-Net can simultaneously model two signals, it is natural to extend it for speaker separation task. The merge branch is removed as two outputs are needed. Permutation invariant training \cite{yu2017permutation} is employed during training to avoid the permutation problem. We conduct the two-speaker separation experiment based on the TIMIT corpus. The batch size is set to 16. For comparison, we train a non-causal version of Conv-TasNet \cite{luo2019conv}, the state-of-the-art method, using the released code\footnote{https://github.com/kaituoxu/Conv-TasNet}.

The results are shown in Table \ref{table4}. We use SDR improvement (SDRi) and PESQ for evaluation. Our method achieves a considerable gain on PESQ by 0.36 and SDRi by 0.82 dB, compared with Conv-TasNet. This suggests that our method is not limited to specific tasks and has the potential to extract different additive parts from a mixture signal.

\section{Conclusion}

\noindent We propose a novel two-branch convolutional neural network to interactively modeling speech and noises for speech enhancement. Particularly, an interaction between two branches is proposed to leverage information learned from the other branch to enhance the target signal modeling. This interaction makes the simultaneous modeling of two signals feasible and effective. Moreover, we design a sophisticated RA block for feature extraction of both branches, which can accommodate the diversities across speech and various noise signals. Evaluations verify the effectiveness of these modules and our method significantly outperforms the state-of-the-art. The two-signal simultaneous modeling paradigm makes it applicable to speaker separation as well. 

\bibliography{snnet}
\end{document}